# Red shift in spectra of galaxies as a consequence of gravitational radiation of the same level as electromagnetic

**S.I. Fisenko, I.S. Fisenko**
"Rusthermosinthes" JSC,
6 Gasheka Str., 12th Storey,
Moscow 125047
Phone: (+7) 956-82-46, Fax: (+7) 956-82-47
E-mail: StanislavFisenko@yandex.ru

**ABSTRACT**

Observational data show that red shift doesn't depend on frequency. The fractional frequency change $z = (n_0-n)/n_0$ is the same for all emission frequencies not only in optical but also in radiorange of a given source ($n_0$ is the frequency of the radiation line of the source, n is the frequency of the same line registered by the detector; $n<n_0$). Such frequency change is a typical feature of Doppler shift, and this suggests the interpretation of red shift as a result of proper motion of galaxies. As for cosmological red shift, which is generally explained by space extension, there is a considerable amount of observational data that don't comply with the theory of purely cosmological origin of red shifts in the spectra galaxies and quasars. Such data are provided, for example, in [1]. In elaboration of the results presented in [2] the red shift is also regarded in this investigation as a widening of electromagnetic radiation spectra, determined by the existence of gravitational radiation of a banded spectrum of the same level as electromagnetic.

## 1. Red Shift as a Physical Phenomenon

Cosmological (metagalactic) red shift is a decrease in the frequency of radiation, observable for all distant sources. Cosmological red shift is often associated with Doppler effect. But in fact Doppler effect has nothing to do with cosmological red shift, which is actually determined by space extension according to GR. There is a theory that the observed red shift of galaxies is the result not only of cosmological red shift due to space extension, but also of red and violet shift of Doppler effect due to proper motion of galaxies. However, the contribution of the cosmological red shift prevails at large distances. The formation of cosmological red shift appears to go on as follows. Take a look at light – an electromagnetic wave running from a distant galaxy. While the light goes through the universe, space extends. The wave package extends alongside space, and the wavelength changes accordingly. If space undergoes double extension, the wavelength and the wave package double as well. This theoretically streamlined image fails in many cases to correspond to observational data. For example, Hubble law is inadequate or doesn't hold at all for objects at a distance less than 10-15 light years, that is exactly for those galaxies the distances to which are most reliably estimated without red shift.

Hubble law doesn't hold good also for those galaxies at very large distances, such as billions of light years, to which the value z > 1 corresponds. The distances to the objects with such a high red shift lose their single-valuedness. At such distances the specific cosmological effects – nonstationarity and space-time curvature – manifest themselves. In particular, the idea of unique, single-valued time becomes nonapplicable (one of the distances – the red shift distance – reaches here $r = v/H = 3,3$ Gpc), as distances depend on the assumed model of the universe and on their reference to a particular moment of time. That is why mere red shift value is used as a characteristic of such large distances (the highest red shift value is registered for object GRB090423 (in Leo), it amounts to z=8,2).

## 2. Banded Spectrum of Gravitational Emission

Generally covariant equations form in gravity relativity theory, as it is well-known, is as follows:

$$R_{ik} - \frac{1}{2} g_{ik} R - \Lambda g_{ik} = \chi T_{ik} \qquad (1)$$

In these equations $\chi$ is the constant, connecting the space-and-time geometric property with the distribution of physical material, so the origin of the equations is not connected with numeric limitation of $\chi$ quantity value. Only the necessity to correspond classical Newtonian gravitation theory brings us to numeric value $\Lambda = 0, \chi = 8\pi G/c^4$, where *G is* Newtonian gravitation constant. The equations with the defined constants are the equations of the Einsteinian general relativity theory (GR). The equations (1) are common mathematical form of gravitational field equations, corresponding to the equivalency principle and general covariance axiom. The equations in form (1) were acquired simultaneously with Einstein but independently from him by Hilbert [3]. In [ 2] there was made a simple and at the same time strict assumption of the existence of such numeric values of gravitational constant К and constant $\Lambda$ in quantum sphere, that bring to stationary states in proper gravitational field, and these are already emitters of gravitational field with Newtonian gravitational constant. The very numeric values of К and $\Lambda$ are estimated independently, exactly within this approach. Herewith we make reference to A. Salam [5], as he was one of the first to pay attention to the fact that Newtonian gravitational constant's numeric value does not conform to quantum level. He was the one to propose the concept of "strong" gravity, that was based on the assumption of f-mesons spin 2 existence, that form SU(3)- multiplet (described by Paul-Firz). It was proven that a possibility of a different link constant along with Newtonian one does not contradict the observed effects [5]. Due to a number of reasons this approach was not developed further. As it is clear now, this numeric value of "strong" gravity constant is to be used in equations (1) with $\Lambda \neq 0$. Besides, precisely with $\Lambda \neq 0$ stationary solutions of general Einsteinian equations can be found, which was noticed by Einstein himself, but after the discovery of functionary solutions with $\Lambda = 0$ by A. Friedman [6], the modern shape of the GRT was finally formed. The decisive argument of GRT to equal $\Lambda$- element to zero is the necessity of right limit passage to Newtonian gravitation theory.

In the simplest (from the point of view of the original mathematical estimations) approach the problem on steady state in proper gravitational field (with constants K and $\Lambda$) is solved in [2]. From the solution of this problem it becomes evident:

a) With numeric values of $K \approx 5.1\times10^{31}$ Nm$^2$kg$^{-2}$ и $\Lambda = 4.4\times10^{29}$ m$^{-2}$ there is a spectrum of electron's stationary states in proper gravitational field (0.511 MeV …0.681 MeV). The main state is detected electron's rest energy 0.511 MeV. *Existence of such numerical value $\Lambda$ denotes a phenomenon having a deep physical sense: introduction into density of the Lagrange function of a constant member independent on a state of the field. This means that the time-space has an inherent curving which is connected with neither the matter nor the gravitational waves.*

b) These stationary states are the emitters of gravitational field with *G* constant.

c) The transition to stationary states in proper gravitational field causes gravitational emission that is characterized by constant K, and with this is the emission of the same level with electromagnetic (electric charge e, gravitational charge $m\sqrt{K}$). In this respect it is meaningless to say that gravitational effects in quantum area are characterized by G constant, this constant belongs only to microscopic level and it cannot be transferred to quantum level (which is evident from negative results of gravitational waves with G constant registration tests, these do not exist).

d) The existence of stationary states in proper gravitational field also completely corresponds to the special relativity theory. According to SRT, relativistic link between energy and impulse is broken, if we assume that full electron's energy is defined only by Lawrence's electromagnetic energy [7]. If we expand the situation it is as follows [7]. Energy and impulse of the moving electron (with the assumption that the distribution of the electric charge is spherically symmetrical) is defined by the expression:

$$\overline{P} = \overline{V} \frac{\frac{4}{3}E_0 / c^2}{\sqrt{1-\beta^2}} \quad (2)$$

$$E = \frac{E_0(1+\frac{1}{3}V^2/c^2)}{\sqrt{1-\beta^2}} \quad (3)$$

If these expressions were at the same time defining full impulse and full energy, the following relator would take place:

$$E = \int (\overline{V} \frac{d\overline{P}}{dt}) dt \quad (4)$$

However this relator cannot take place as the integral in the right part equals to:

$$\frac{\frac{4}{3}E_0}{\sqrt{1-\beta^2}} + const \quad (5)$$

If we find that impulse contrary to the energy has a strictly electromagnetic character, then to $E^{'}$ of the moving and full energy $E_0^{'}$ of the resting electron, and also to rest mass $E_0$, the following relators will take place:

$$E' = \frac{E_0^{'}}{\sqrt{1-\beta^2}}, \quad E_0^{'} = \frac{4}{3}E_0, \quad m_0 = \frac{E_0^{'}}{c^2} = \frac{4}{3}\frac{E_0}{c^2}, \tag{6}$$

Where rest mass $m_0$ is defined by the following expression:

$$\overline{P} = \frac{m_0 \overline{V}}{\sqrt{1-\beta^2}} \tag{7}$$

Then from (6) it follows that full energy of resting electron equals to $\frac{4}{3}$ of its Lawrence's electromagnetic energy. Numeric data of the electron's stationary states spectrum in proper gravitational field fully correspond to it.

e) The spectrum of electron's stationary states in proper gravitational field and transitions to stationary states are represented on Fig. 1. We should mention straight away that the numeric value is approximate. The largest inaccuracy belongs to the numeric value of the first stationary state $E_1$, but it is more and more accurate coming closer to $E_\infty = 171 keV$.

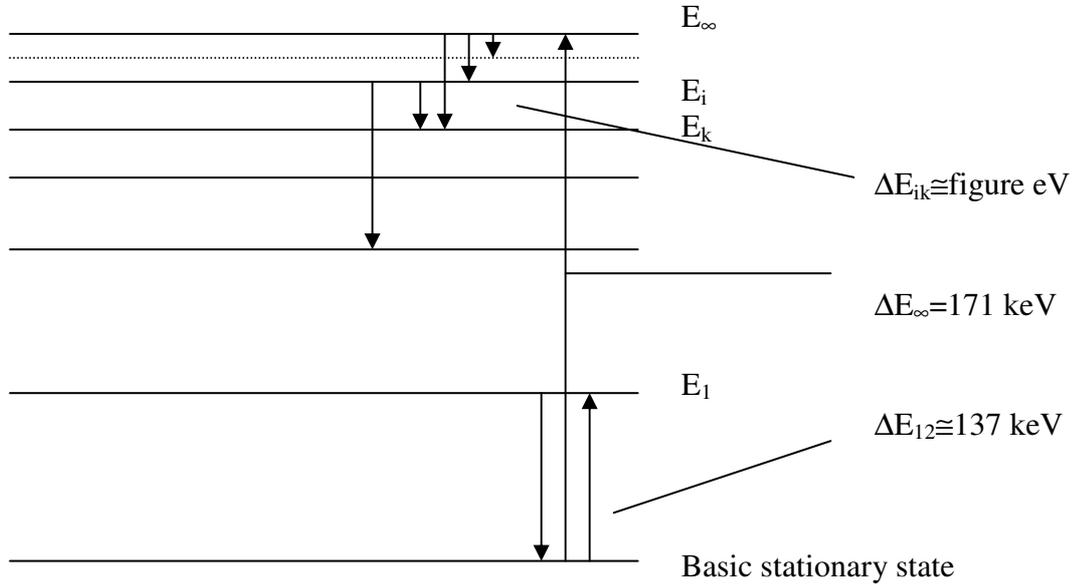

Figure 1. Transition over stationary states of electron in proper gravitational.

### 3. Spectrum Resonances of Stationary States of Electromagnetic and Gravitational Interactions

Fig. 2 shows electronic energy levels of multielectron atoms, while fig. 3 presents rotational level in the nucleus of $^{171}$Er (as an element taken approximately from the middle of the Periodic table). Simple comparison of these spectra with the spectrum given in fig. 1 shows the possibility of resonance transitions between these energy levels. The result of such resonance transitions is additional widening of corresponding spectrum lines, which is experimentally proved. Fig.4 shows characteristic parts of micropinch soft X-ray radiation spectrum. Micropinch spectrum line widening does not correspond to existing electromagnetic conceptions but corresponds to such plasma thermodynamic states which can only be obtained with the help of compression by gravitational field, the radiation flashes of which take place during plasma thermalization in a discharge local space. Such statement is based on the comparison of experimental and expected parts of the spectrum shown in Fig. 4. Adjustment of the expected spectrum portion to the experimental one [8] was made by selecting average values of density ρ, electron temperature $T_e$ and velocity gradient U of the substance hydrodynamic motion.

As a mechanism of spectrum lines widening, a Doppler, radiation and impact widening were considered. Such adjustment according to said widening mechanisms does not lead to complete reproduction of the registered part of the micropinch radiation spectrum. This is the evidence (under the condition of independent conformation of the macroscopic parameters adjustment) of the existence of an additional widening mechanism due to electron excited states and corresponding gravitational radiation spectrum part already not having clearly expressed lines because of energy transfer in the spectrum to the long-wave area.

That is to say that the additional mechanism of spectral lines widening of the characteristic electromagnetic radiation of multiple-charge ions (in the conditions of plasma compression by radiated gravitational field) is the only and unequivocal way of quenching electrons' excited states at the radiating energy levels of ions and exciting these levels by gravitational radiation at resonance frequencies. *Such increase in probability of ion transitions in other states results in additional spectral lines widening of the characteristic radiation.* The reason for quick degradation of micropinches in various pulse high-currency discharges with multiple-charge ions is also clear. There is only partial thermalization of accelerated plasma with the power of gravitational radiation not sufficient for maintaining steady states. Thus, resonance transitions determined by spectral foldover, electrons and multiple-charge ions in the situation under consideration, bring about the widening of radiation spectrum of dense high-temperature plasma (as it is exactly in such plasma that the conditions of resonance transitions are realized).

The qualitatively same situation takes place in the plasma of space objects, differing in quantitatively in the range of spectrum widening.

A contribution into such widening of spectral lines may be given by a resonance of the energy spectra of electrons (in proper gravitational field) with the energy spectra of multicharged ions (see fig. 2) and the spectrum of rotational energy levels of nuclei (see fig. 3). This widening of spectral radiation lines of space objects must take place in the whole radiation spectrum of these objects.

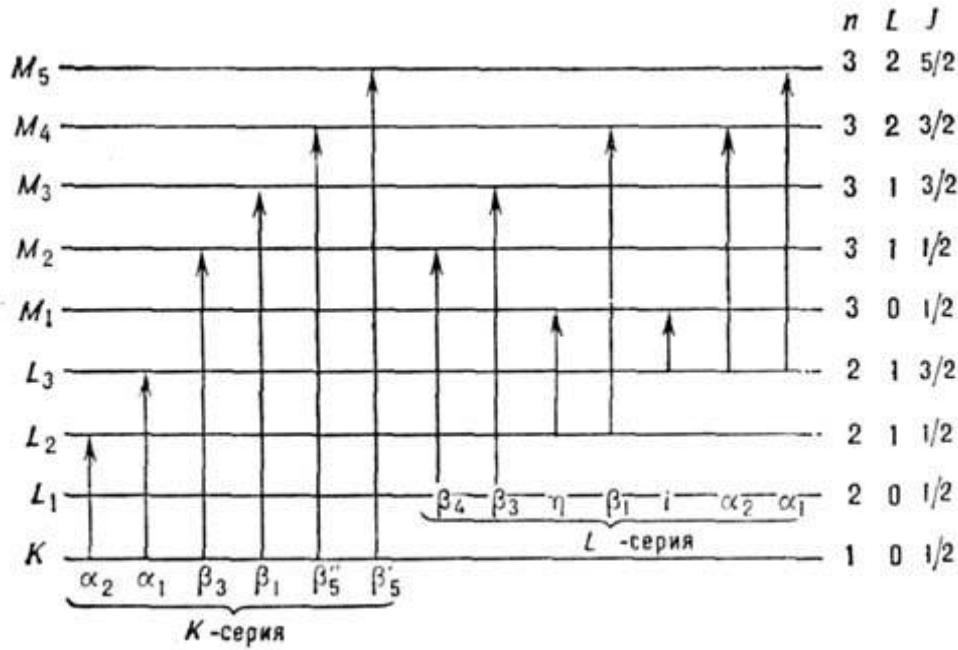

Fig. 2. A scheme of K-, L- and M-levels of energy of the atom, and the main lines of K- and L- series; n, l, j are the principal, the orbital and the inner quantum numbers of energy levels к, $L_1$, $L_2$ etc. The energies of photons of the main lines reach units and scores of keV.

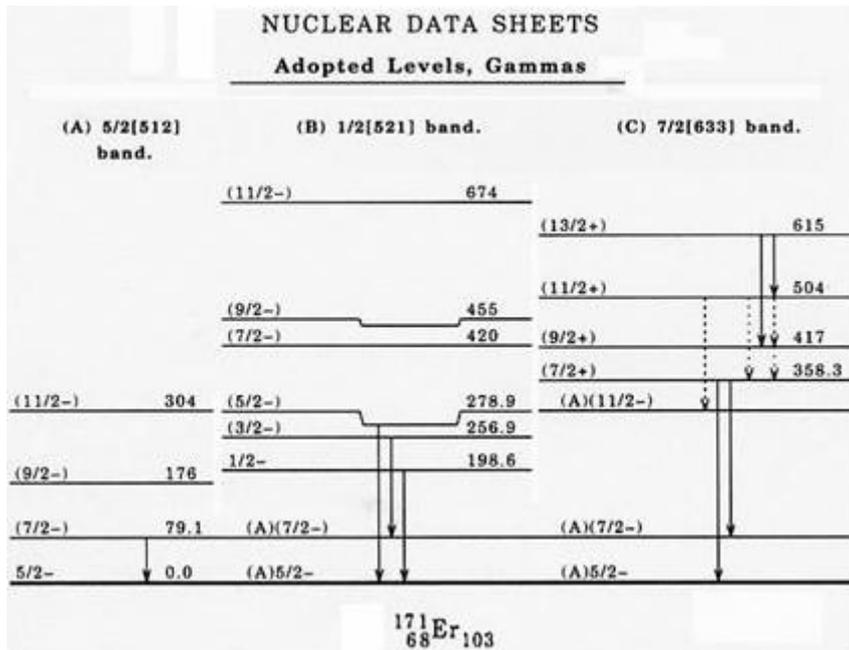

Fig. 3. Regular rotational bands in the nucleus of $^{171}$Er. Lower rotational energy levels of nuclei are apart from the main one by scores and hundreds of keV.

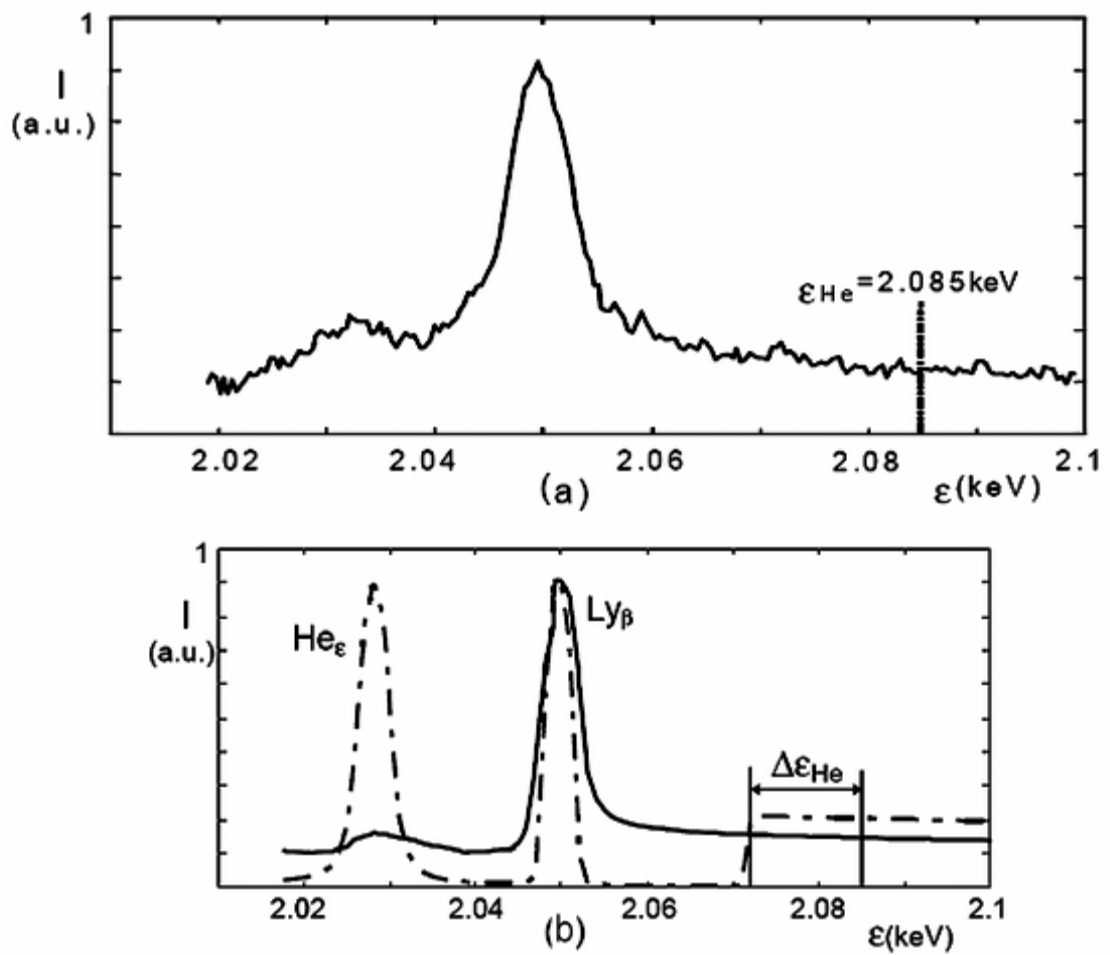

Fig. 4. The experimental and the calculated areas of a micropinch spectrum, normalized for line Ly$\beta$ intensity, in the area of the basic state ionization threshold of He-like elements. The firm line in variant b) corresponds to density of 0,1 g/cm$^3$, and the dotted line – to that of 0,01 g/cm$^3$; it was assumed that $T_e$=0,35 keV, [8].

As is known, relict radiation is only a part of the general cosmic-ray background, the full spectrum of which is displayed in fig. 5 [9].

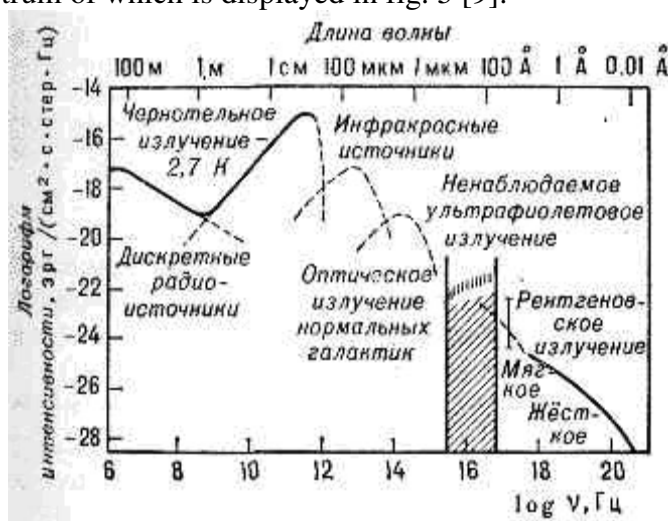

Fig. 5. The spectrum of background electromagnetic radiation of the Universe. The solid line indicates observational data, the dotted line shows theoretical estimates.

[9] provides a general characteristic of this spectrum, that is:

Long wave radioemission ($\nu < 600$ Hz; $\lambda > 50$ cm). Radiotelescopes receive cosmic background radiation as well as synchrotron radiation of relativistic electrons in the interstellar matter of Galaxy, which makes it difficult to distinguish the background radiation. The synchrotron radiation is distributed on the sky very nonuniformly. Of interest is an area in the sky with the minimal radiance temperature reaching 80 K at a frequency of 178 MHz. Apparently, this is the upper radiance temperature limit of background radiation at this frequency. The extragalactic component can be distinguished only if the Galaxy radiation spectrum differs from the spectrum of background radiation. However, they are quite close. A careful study shows that background radiation radiance temperature at a frequency of 178 MHz is close to 30 K, while the spectral index coincides with the average radiogalaxy emission spectral index $\alpha \approx 0,75$. This makes it possible to distinguish the background radiance temperature at any wavelength in metric band

$$T_b \approx 30(\lambda/1,7\text{м})^{2,75}\text{K}, \quad I_\nu = 3 \cdot 10^{-19} \times (\lambda/1,7\text{м})^{0,75} \text{эрг}/(\text{см}^2 \cdot \text{с} \cdot \text{Гц} \cdot \text{ср})$$

The coincidence of background and radiogalaxy spectral indices leads to the suggestion that long-wave background radiation is a combined radiation by distant high-energy discrete radioemission sources: radiogalaxies and quasars. However, the space density of radiogalaxies and their radioemission observed in the vicinity of our Galaxy have proved insufficient for the explanation of background radiation. The resolution of this issue was made possible only after a careful calculation of low-energy (and consequently, distant) sources of radioemission. Distant high-energy radiogalaxies and quasar are observed now as low-energy sources. It is these numerous sources that have proved to determine background radiation in the long-wave area.

Microwave background radiation ($6 \cdot 10^8 \text{Гц} < \nu < 10^{12}\text{Гц}; \ 300\text{мкм} < \lambda < 50$ cm). Calculations in centimeter and millimeter wavelenth areas , which were carried out in 1965, lead to the discovery of isotropic emission with spectrum of an ideal black body and the temperature $\approx 2,7$ K. This radiation makes the main contribution into the energy density and photon concentration of background radiation.

Infrared diapason ($10^{12}\text{Гц} < \nu < 3 \cdot 10^{14}\text{Гц}; \ 1\text{мкм} < \lambda < 300$ μm). This spectral interval has only the upper background radiation intensity limits determined. In fact, observations in the infrared band are scarce, as they are largely hampered by molecular absorption and emission in upper atmosphere. Surface measurements in atmospheric transparency windows are only possible with при $\lambda$ <25 μm. Observations of celestial objects within the interval of 25 μm<$\lambda$ <300 μm are carried out from rockets, balloons and high-altitude airplanes. The development of observation technology has lead to the discovery of infrared excess in the spectra of many discrete sources. A considerable number of galactic objects, including certain star types and an array of planetary and "infrared" nebulae, have turned out to be anomalously bright in the short-range ($\lambda$<25 μm) infrared diapason. In the majority of cases these are cold stars (condensing protostars and giant stars) with the temperature < 2000 K, or dust complexes, that reemit the ultraviolet and optical radiation of hot stars situated within them. But the emittance of all these objects is not very high, that is why the total radiation of sources of this type in other galaxies cannot determine the main contribution into background radiance.

Optical diapason ($3 \cdot 10^{14}$ Гц $< \nu < 10^{15}$ Гц; $3000$Å$ < \lambda < 1$ мкм). Surface observations have not yielded any traces of the isotropic optical component of background radiation. The upper limit has turned out to be about 100 times lower than full observed sky radiance. With a knowledge of radiation spectra of separate galaxies, their integral emission can be found. Besides it appears that the main contribution into optical background radiation is made by normal galaxies (to be precise the radiation of stars belonging to them).

Ultraviolet diapason.
This area of the spectrum falls into two parts: one of them is accessible for observations from satellites and rockets, the other is fundamentally inaccessible for direct observations from Solar system.

a) The observable diapason ($10^{15}$ Гц $< \nu < 3,3 \cdot 10^{15}$ Гц; $912$Å$ < \lambda < 3000$Å). Sky radiance in the ultraviolet area of the spectrum is determined by the radiation of hot stars of our Galaxy. Apparently, the higher temperature T of the star surface, the more photons in ultraviolet radiation it emits. However, the number of stars having this temperature sharply decreases with the growth of T. The integral radiation of Galaxy stars drops as rapidly with the decrease of the wave-length. That is why it was expected that it would be easier to distinguish the extragalactic component within the ultraviolet range, than in the optical one, and that it would carry the information mainly about nonstellar sources – galaxy nuclei, quasars and interstellar gas. It is true that the observable ultraviolet range also encompasses a powerful radiation, determined by reemission by interplanetary hydrogen of L$\alpha$ line of solar origin. However, this radiation may be excluded by filters. Despite all the efforts the metagalactic ultraviolet radiation has not been distinguished yet. Only the upper limits of its intensity have been experimentally determined.

b) The unobservable diapason ($3,3 \cdot 10^{15}$ Гц $< \nu < 3 \cdot 10^{16}$ Гц; $100$Å$ < \lambda < 912$Å). This spectrum area is fundamentally inaccessible for direct observations within Solar system because of absorption of ultraviolet radiation photons by neutral interstellar hydrogen. There is only an indirect method of intensity assessment of the ionizing background radiation. Background radiation must create hydrogene ionization zones around galaxies. If the background level were too high, the ultraviolet range photons would be able to ionize the interstellar gas. In reality radio observations in hydrogen line of 25 cm have lead to the discovery of neutral gas far off the optical borders of galaxies. Hydrogen density is extremely low there, and the fact that it is not ionized ther gives grounds to speak about low intensity of ultraviolet background radiation; its upper limit is 100 times lower that in the adjoining observable diapason. Hydrogen in the galaxy periphery has turned out to be a 100 more sensitive detector than the detectors on satellites and rockets. The obtained limit is not very low: it corresponds to 10000 ionizing photons falling on 1 cm$^2$ of galaxies surface in 1 second.

X-ray diapason ($3 \cdot 10^{16}$ Гц $< \nu < 10^{20}$ Гц; $0{,}1\text{Å} < \lambda < 100\text{Å}$).

Rocket, satellite and balloon investigations have shown that radiation in the classic X-ray area ($\lambda \sim 1\text{-}10\text{Å}$) is highly isotropic, i.e. has an extragalactic nature. Only in the area of soft X-rays (for photons with energy $\varepsilon > 250$ eV) a strong dependence of diffuse radiation intensity on galactic coordinates is observed.

The main sources of background X-ray radiation are still not known with certainty. It appears that these are the nuclei of galaxies, hot intergalactic gas in galactic clusters, and quasars. However, the radiation of these objects cannot provide for more than 50% of background radiation intensity in the X-ray range. The origin of X-ray background radiation may be connected with the scattering of low-frequency photons by relativistic electrons of cosmic rays. In the process of such scattering the photon energy increases manifold and they get into the X-ray diapason. In galaxy nuclei comptonization is apparently effective, which lead to the formation of hard X-rays in hot nonrelativistic Maxwellian plasma. Another crucially important mechanism of X-ray photon radiation is deceleration radiation of hot gas.

Fig. 1 above displays the energy spectrum of stationary states of an electron in proper gravitational field, in its simplest form. The account of all quantum states will yield a more detailed spectrum, especially in its upper part. This follows directly from the approximation in which the spectrum given in fig. 1 was found. But already this approximation, with comparison of spectra given in fig. 2 and fig. 3, provides for the possibility of resonance transitions between the stationary state levels of electromagnetic and gravitational interactions. The same transitions will take place in case of typical atomic-molecular spectra, such as the ones given in fig. 6 and fig. 7.

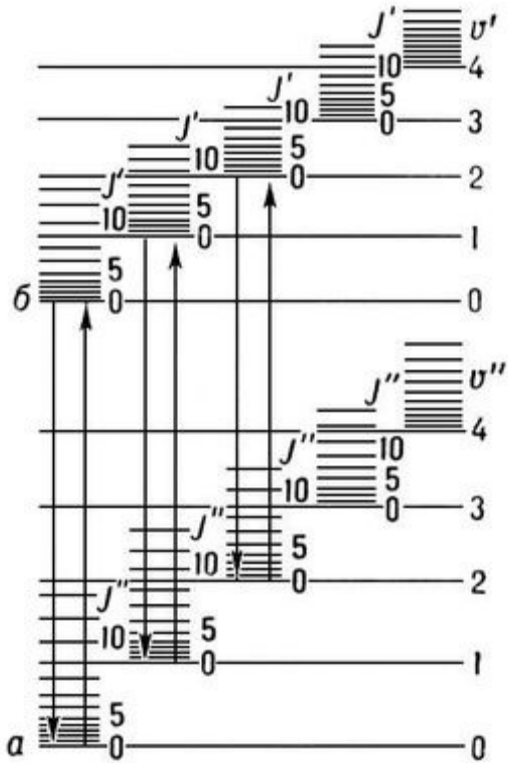

Fig. 6. The scheme of energy levels of a binary molecule: a and b are electronic levels; $v'$ and $v''$ are quantum numbers of vibrational levels; $J'$ and $J''$ are quantum numbers of rotational levels.

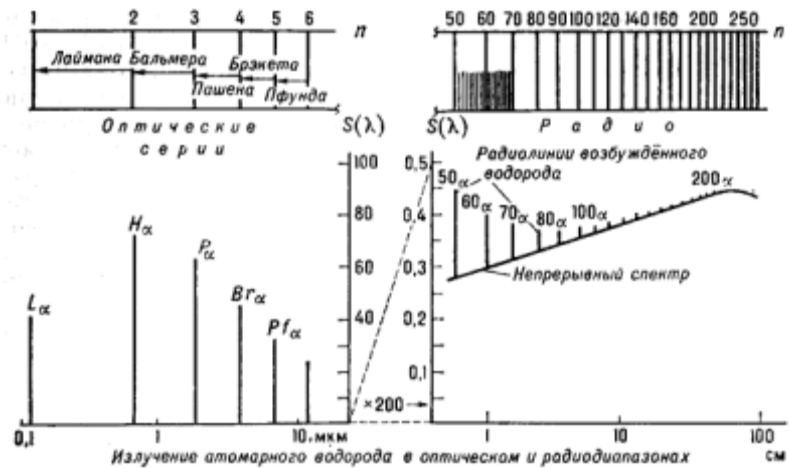

Fig. 7. Spectral lines of hydrogen in optical and radio diapason. Numeration refers to the final level of the atom after the emission of a quantum of a corresponding frequency; $\alpha$ is the first member of a corresponding series, the line intensity $S(\lambda)$ is given in random units.

Thus, in qualitative respect a part of radiation should be present in the whole spectrum of electromagnetic, as well as of local sources of diffuse radiation, as a result of resonance transitions between the spectra of electromagnetic and gravitational interactions. This means that in dense high-temperature plasma on multicharged ions (momentum high-current discharges), as well as in the plasma of space objects, the presence of excited states of electrons (as a result of undeniably existing spectrum of stationary states in proper gravitational field), regardless of further development of the situation, will lead to widening of the corresponding electromagnetic radiation spectrum lines. This is exactly what is observed in laboratory experiments with plasma (see fig. 4), and in the shift of the lines of galaxy radiation spectrum (alongside Doppler shift). The further development of the situation is determined by the parameters of plasma.

The populations of quantum levels and, consequently, the spectrum characteristics appear considerably different in cases of plasma with different amount of multicharge ions. Phisically it is determined by the competition of processes of radiative transition (i.e. spontaneous emission) and non-radiative transition in case of a collision of an atom with an electron. In case of excitation of upper energy levels of an electron in multicharged ion plasma (in the process of drag in ion nuclei), cascade transitions to lower energy levels will bring about the transfer of gravitational radiation into long-wave part of the spectrum, with the following blocking and intensification of radiation [2]. In case when the consentration of multicharged ions is insignificant and their energy states spectrum does not allow to quench the lower excited state of electrons, a micropinch [8] will take place, followed by its rapid decomposition. This scheme is specially characteristic of laboratory plasma of multicharged ions, but not of the plasma of cosmic objects. The colossal geometrical dimensions of cosmic objects plasma will naturally cause absorption of the emitted gravitational quanta, leaving its impact in widened spectra of electromagnetic radiation.